\PassOptionsToPackage{unicode}{hyperref}
\PassOptionsToPackage{hyphens}{url}
\PassOptionsToPackage{dvipsnames,svgnames,x11names,pdftex}{xcolor}
\documentclass[
  12pt]{article}

\usepackage{amsmath,amssymb}
\usepackage{iftex}
\ifPDFTeX
  \usepackage[T1]{fontenc}
  \usepackage[utf8]{inputenc}
  \usepackage{textcomp} 
\else 
  \usepackage{unicode-math}
  \defaultfontfeatures{Scale=MatchLowercase}
  \defaultfontfeatures[\rmfamily]{Ligatures=TeX,Scale=1}
\fi
\usepackage{lmodern}
\ifPDFTeX\else  
\fi
\IfFileExists{upquote.sty}{\usepackage{upquote}}{}
\IfFileExists{microtype.sty}{
  \usepackage[]{microtype}
  \UseMicrotypeSet[protrusion]{basicmath} 
}{}
\makeatletter
\@ifundefined{KOMAClassName}{
  \IfFileExists{parskip.sty}{%
    \usepackage{parskip}
  }{
    \setlength{\parindent}{0pt}
    \setlength{\parskip}{6pt plus 2pt minus 1pt}}
}{
  \KOMAoptions{parskip=half}}
\makeatother
\usepackage{xcolor}
\makeatletter
\ifx\paragraph\undefined\else
  \let\oldparagraph\paragraph
  \renewcommand{\paragraph}{
    \@ifstar
      \xxxParagraphStar
      \xxxParagraphNoStar
  }
  \newcommand{\xxxParagraphStar}[1]{\oldparagraph*{#1}\mbox{}}
  \newcommand{\xxxParagraphNoStar}[1]{\oldparagraph{#1}\mbox{}}
\fi
\ifx\subparagraph\undefined\else
  \let\oldsubparagraph\subparagraph
  \renewcommand{\subparagraph}{
    \@ifstar
      \xxxSubParagraphStar
      \xxxSubParagraphNoStar
  }
  \newcommand{\xxxSubParagraphStar}[1]{\oldsubparagraph*{#1}\mbox{}}
  \newcommand{\xxxSubParagraphNoStar}[1]{\oldsubparagraph{#1}\mbox{}}
\fi
\makeatother

\usepackage{longtable,booktabs,array}
\usepackage{calc} 
\usepackage{etoolbox}
\makeatletter
\patchcmd\longtable{\par}{\if@noskipsec\mbox{}\fi\par}{}{}
\makeatother
\IfFileExists{footnotehyper.sty}{\usepackage{footnotehyper}}{\usepackage{footnote}}
\makesavenoteenv{longtable}
\usepackage{graphicx}
\makeatletter
\def\maxwidth{\ifdim\Gin@nat@width>\linewidth\linewidth\else\Gin@nat@width\fi}
\def\maxheight{\ifdim\Gin@nat@height>\textheight\textheight\else\Gin@nat@height\fi}
\makeatother
\setkeys{Gin}{width=\maxwidth,height=\maxheight,keepaspectratio}
\makeatletter
\def\fps@figure{htbp}
\makeatother

\makeatletter
\@ifpackageloaded{caption}{}{\usepackage{caption}}
\AtBeginDocument{%
\ifdefined\contentsname
  \renewcommand*\contentsname{Table of contents}
\else
  \newcommand\contentsname{Table of contents}
\fi
\ifdefined\listfigurename
  \renewcommand*\listfigurename{List of Figures}
\else
  \newcommand\listfigurename{List of Figures}
\fi
\ifdefined\listtablename
  \renewcommand*\listtablename{List of Tables}
\else
  \newcommand\listtablename{List of Tables}
\fi
\ifdefined\figurename
  \renewcommand*\figurename{Figure}
\else
  \newcommand\figurename{Figure}
\fi
\ifdefined\tablename
  \renewcommand*\tablename{Table}
\else
  \newcommand\tablename{Table}
\fi
}
\@ifpackageloaded{float}{}{\usepackage{float}}
\floatstyle{ruled}
\@ifundefined{c@chapter}{\newfloat{codelisting}{h}{lop}}{\newfloat{codelisting}{h}{lop}[chapter]}
\floatname{codelisting}{Listing}

\makeatother
\makeatletter
\@ifpackageloaded{caption}{}{\usepackage{caption}}
\@ifpackageloaded{subcaption}{}{\usepackage{subcaption}}
\makeatother

\ifLuaTeX
  \usepackage{selnolig}  
\fi
\usepackage[]{natbib}
\usepackage{bookmark}

\IfFileExists{xurl.sty}{\usepackage{xurl}}{} 
\hypersetup{
  pdftitle={Title},
  pdfauthor={Author 1; Author 2},
  pdfkeywords={3 to 6 keywords, that do not appear in the title},
  colorlinks=true,
  linkcolor={blue},
  filecolor={Maroon},
  citecolor={Blue},
  urlcolor={Blue},
  pdfcreator={LaTeX via pandoc}}
  
\usepackage{amsthm}
\usepackage{amsmath}
\usepackage{accents}
\numberwithin{equation}{section}
\usepackage{bm}
\usepackage{mathtools,mathrsfs}
\usepackage{graphicx}
\usepackage{float}
\makeatletter
\def\wb{\accentset{{\cc@style\underline{\mskip10mu}}}} 
\makeatother

\usepackage{booktabs}
\usepackage{tabularx}
\usepackage{multirow}
\usepackage{etoolbox}
\usepackage{xcolor}

\theoremstyle{plain}

\newtheorem{theorem}{Theorem}[section]
\newtheorem{corollary}{Corollary}[section]
\theoremstyle{definition}
\newtheorem{definition}{Definition}

\newtheorem{assumption}{Assumption}
\newtheorem{proposition}{Proposition}[section]

\usepackage{algorithm}
\usepackage{algpseudocode}

\usepackage{booktabs}
\usepackage{tabularx}
\usepackage{multirow}
\usepackage{etoolbox}

\numberwithin{equation}{section}
\usepackage{bm}
\usepackage{mathtools,mathrsfs}
\usepackage{comment}
\usepackage{graphicx}
\usepackage{float}
\usepackage{fix-cm}

\newcommand{\wh}{\widehat}
\newcommand{\wt}{\widetilde}
\newcommand{\mr}{\mathring}

\newcommand{\cA}{\mathcal{A}}
\newcommand{\bA}{\boldsymbol{A}}
\newcommand{\cB}{\mathcal{B}}
\newcommand{\bB}{\boldsymbol{B}}
\newcommand{\bb}{\boldsymbol{b}}
\newcommand{\cC}{\mathcal{C}}
\newcommand{\bc}{\boldsymbol{c}}

\newcommand{\bG}{\boldsymbol{G}}
\newcommand{\bg}{\boldsymbol{g}}

\newcommand{\bbI}{\mathbb{I}}

\newcommand{\rK}{\mathring{K}}
\newcommand{\dK}{\ddot{K}}
\newcommand{\cL}{\mathcal{L}}
\newcommand{\bO}{\boldsymbol{O}}

\newcommand{\cP}{\mathcal{P}}

\newcommand{\bR}{\boldsymbol{R}}
\newcommand{\bS}{\boldsymbol{S}}
\newcommand{\bU}{\boldsymbol{U}}
\newcommand{\bV}{\boldsymbol{V}}

\newcommand{\bW}{\boldsymbol{W}}

\newcommand{\bZ}{\boldsymbol{Z}}

\newcommand{\bpi}{\boldsymbol{\pi}}

\newcommand{\bGamma}{\boldsymbol{\varGamma}}
\newcommand{\bLambda}{\boldsymbol{\varLambda}}
\newcommand{\brho}{\boldsymbol{\rho}}

\newcommand{\balpha}{\boldsymbol{\alpha}}

\newcommand{\cX}{\mathcal{X}}
\newcommand{\bX}{\boldsymbol{X}}

\newcommand{\bbE}{\mathbb{E}}
\newcommand{\bbM}{\mathbb{M}}
\newcommand{\bbO}{\mathbb{O}}
\newcommand{\bbP}{\mathbb{P}}
\newcommand{\bbR}{\mathbb{R}}

\renewcommand{\|}{\Vert}
\renewcommand{\d}{\operatorname{d}}
\newcommand{\Reg}{\operatorname{Reg}}
\newcommand{\err}{\operatorname{err}}
\newcommand{\diag}{\operatorname{diag}}

\usepackage{caption}
\newcommand{\anon}{1}

\begin{document}

\def\spacingset#1{\renewcommand{\baselinestretch}%
{#1}\small\normalsize} \spacingset{1}


\if1\anon
{
  \title{\bf Sparsity-Corrected Tensor Clustering for Multi-Layer Networks}
  \author{Xuefei Wang \\
    School of Statistics and Data Science, Southeast University\\
    and \\
    Gaorong Li\\
    School of Statistics, Beijing Normal University\\
    and \\
    Wenyang Zhang \thanks{The corresponding author, Email: \texttt{wenyangzhang@um.edu.mo}.}\\
    Faculty of Business Administration \& APAEM, University of Macau
    }
    \date{}
  \maketitle
} \fi

\if0\anon
{
  \bigskip
  \bigskip
  \bigskip
  \begin{center}
    {\LARGE\bf Sparsity-Corrected Tensor Clustering for Multi-Layer Networks}
\end{center}
  \medskip
} \fi

\bigskip
\begin{abstract}
Multi-layer networks provide complementary views of interactions among the same set of nodes. In practice, layers with similar community structures may have substantially different network sparsity, causing sparsity variation to obscure the structural patterns of interest. To address this problem, a sparsity heterogeneous mixture multi-layer stochastic block model is proposed, which separates layer-specific sparsity from group-specific community structure. Based on the induced low-rank tensor representation, a sparsity-corrected tensor clustering method is proposed to remove the layer-specific scale effect before clustering, without explicitly estimating the sparsity parameters. Theoretical guarantees for tensor decomposition, layer clustering, and community detection are established. The results further characterize the range of sparsity heterogeneity allowed for consistent recovery, including regimes where the sparsity ratio diverges with the network dimensions. Numerical studies and an application to international trade networks from BACI demonstrate the effectiveness of the proposed method under heterogeneous layer sparsity.
\end{abstract}

\noindent%

\vfill

\newpage
\spacingset{1.8} 

\section{Introduction}\label{sec:1}

Networks provide a natural representation of relational data, where nodes correspond to entities and edges describe their interactions. In many applications, the same set of nodes is observed through multiple types of relations or under different conditions, giving rise to a multi-layer network. Such data have been widely studied in sociology \citep{gaoLargescaleMultilayerAcademic2023}, genomics \citep{xieJointAnalysisMultiple2023}, neuroscience \citep{wangHeterogeneityAnalysisMultistate2024}, and linguistics \citep{liuQuantitativeLinguisticAnalysis2024}. Compared with a single network, multi-layer networks provide richer information but also introduce variation across layers. An important question is therefore how to combine information from related layers while retaining the structural differences among them.

Let $\mathcal{G}(\mathcal{V},{\mathcal{E}^{(1)},\ldots,\mathcal{E}^{(L)}})$ denote an $L$-layer network on a common node set $\mathcal{V}=[N]$ and edge set of layer $\mathcal{E}^{(l)}$, where $[N]=1,\ldots,N$ and $l \in [L]$. The observations are represented by an adjacency tensor $\cA\in\{0,1\}^{N\times N\times L}$, with $\cA_{ijl}=1$ if nodes $i$ and $j$ are connected in layer $l$, and $\cA_{ijl}=0$ otherwise. In addition to dependence among the layers, two types of variation are often present. Different layers may have different community structures, and layers with similar community structures may have substantially different network sparsity, causing network sparsity variation to obscure the structural patterns of interest.

Several block models have been proposed to describe structural variation across layers. The multi-layer stochastic block model (MLSBM) \citep{valles-catalaMultilayerStochasticBlock2016,barbillonStochasticBlockModels2017} assumes a common community structure while allowing the connection probability matrices to vary across layers. This model has been extensively studied for community detection \citep{paulSpectralMatrixFactorization2020,leiBiasadjustedSpectralClustering2023,wangMomentIntegratedBiasAdjustedSpectral2026}. The random effects stochastic block model \citep{paulRandomEffectsStochastic2020} and dynamic stochastic block models \citep{linDynamicClusteringHeterophilic2024} further allow node memberships to vary. When a large collection of layers contains only a few recurring community structures, it is natural to group the layers. The strata multi-layer stochastic block model \citep{stanleyClusteringNetworkLayers2016} and the mixture multi-layer stochastic block model (MMSBM) \citep{jingCommunityDetectionMixture2021} follow this idea by representing the layers as a mixture of stochastic block models \citep{hollandStochasticBlockmodelsFirst1983}. In particular, these two models allow different latent layer groups to have different node memberships and different numbers of communities. \cite{jingCommunityDetectionMixture2021} further proposed TWIST to recover the layer memberships together with the local and global node communities. Under MMSBM, however, layers belonging to the same group share the same connection probability matrix, and hence have the same sparsity structure.

Layer-specific variation has also been considered under other multi-layer network models. \cite{paulCommunityDetectionMultirelational2016} proposed the restricted multi-layer stochastic block model (RMLSBM), where additive layer effects in a logistic model account for layer-wise sparsity variation. A more recent latent space approach in \cite{kaganLatentSpaceModels2025} further separates shared, group-specific, and network-specific components to describe heterogeneity across networks. These studies address important forms of layer variation, but their inferential targets are different from layer clustering. In particular, \cite{paulCommunityDetectionMultirelational2016} studied community recovery under a common node partition, while \cite{kaganLatentSpaceModels2025} focused on the estimation of latent network embeddings. Moreover, their theoretical analysis does not characterize the allowable layer-wise sparsity heterogeneity through an explicit heterogeneity ratio.

It remains poorly understood how to recover latent layer groups when layers sharing the same community structure may still have different sparsity levels. Similarity in community structure does not necessarily imply similarity in network sparsity. When the two sources of variation are not distinguished, sparsity may obscure the structural patterns of interest and lead to inaccurate layer clustering. It is also unclear how much within-group sparsity heterogeneity can be tolerated for consistent layer clustering, especially when the heterogeneity itself increases with the network dimensions.

\begin{figure}[htbp]
    \centering
    \includegraphics[width=0.9\linewidth]{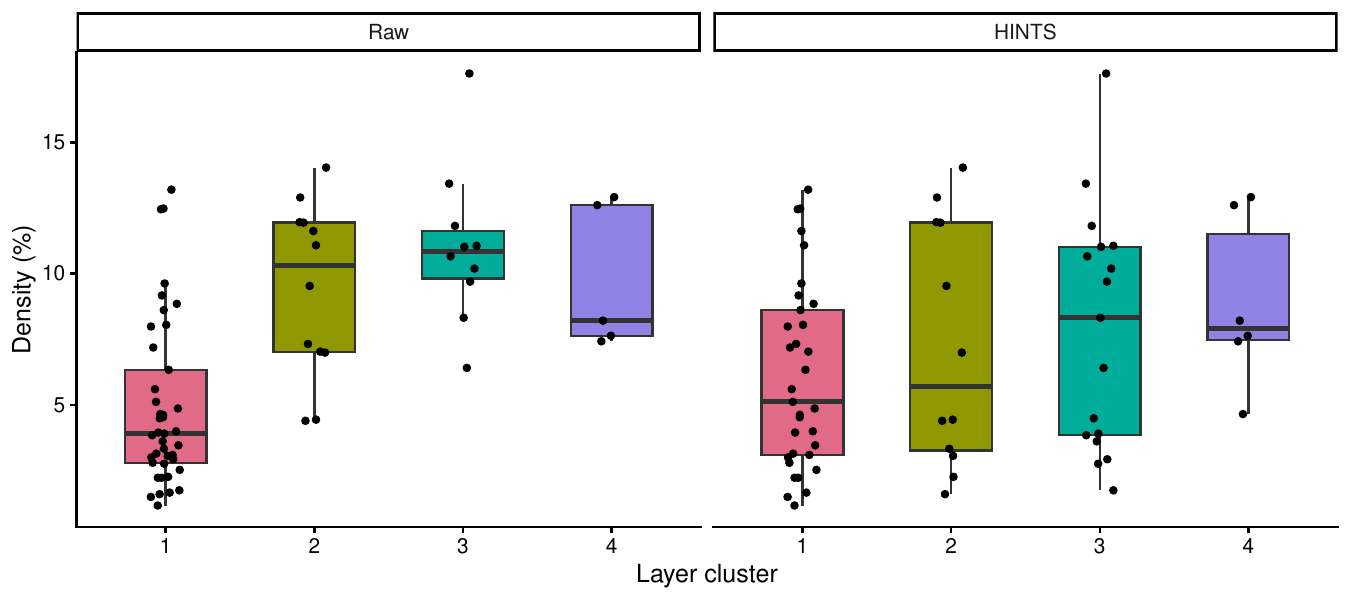}
    \caption{Product clustering from the raw layer embeddings (left) and sparsity-corrected layer embeddings (right). Layer density shown by the vertical position reflects how broadly each product is traded across economy pairs.}
    \label{fig:BACI_density}
\end{figure}

This phenomenon is not merely a modeling concern. For example, consider the BACI trade networks, where each product category defines a network of trading relationships among economies; a detailed description is given in Section~\ref{sec:application}. Figure~\ref{fig:BACI_density} shows how sparsity correction affects product clustering. Under the raw embedding, products traded by relatively few economy pairs tend to be grouped together, even when their international trading structures are quite different. For example, textile materials, ships, railway equipment, and clocks and watches are grouped together largely because their trade networks are relatively sparse. After sparsity correction, ships and railway equipment are grouped with machinery and transport products, clocks and watches with precision products, and basketware with apparel. Thus, products with similarly limited trade prevalence may nevertheless exhibit quite different patterns of trading relationships across economies. The sparsity-corrected clustering reveals these differences rather than grouping products primarily by how broadly they are traded.

This example motivates separating layer-specific sparsity from the underlying connectivity structure. To formalize this distinction, we propose the Sparsity Heterogeneous Mixture Multi-layer Stochastic Block Model (SH-MMSBM). For $l \in [L]$ and $m \in [M]$, if layer $l$ belongs to group $m$, the connection probability matrix is decomposed as $\bB^{(l)}=\rho_l\bB_0^{(m)}$, where $\bB_0^{(m)}$ determines the group-specific community structure and $\rho_l$ controls the sparsity of layer $l$. This formulation allows us to study directly how sparsity heterogeneity affects layer clustering.

\begin{figure}[H]
    \centering
    \includegraphics[width=\linewidth]{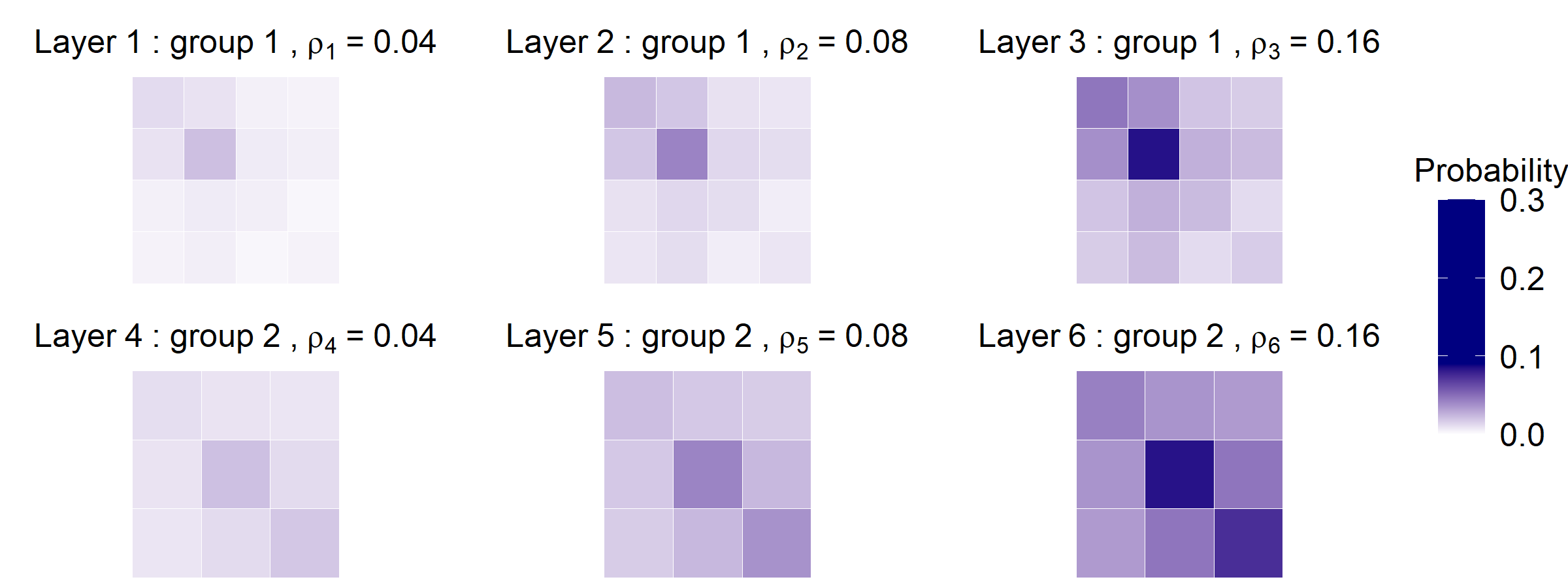}
    \caption{Visualization of the connection probability matrices $\bB^{(l)}$. Layers in the same group share the same connection pattern with heterogeneous sparsity.}
    \label{fig:demo1_34}
\end{figure}

Figure~\ref{fig:demo1_34} gives a controlled illustration of the same phenomenon. The layers are generated from two groups with different community structures, while their sparsity levels vary within each group. In this example, TWIST \citep{jingCommunityDetectionMixture2021} and the adaptive merging procedure in \cite{zhangEfficientEstimationLongitudinal2025} separate layer 6 from the remaining layers rather than recover the two structural groups. The example shows that a layer representation suitable for clustering should distinguish variation in community structure from variation in network sparsity.

Tensor methods provide a natural representation for this purpose because the mode-3 factor describes the relations among layers \citep{zhangTensorSVDStatistical2018}. Under SH-MMSBM, however, the oracle layer embedding contains both pieces of information: the layer group determines its direction and the layer-specific sparsity parameter changes its magnitude. Hence, layers from the same group may be spread along the same direction when their sparsity levels differ. Direct distance-based clustering of the raw layer embedding can then be affected by the variation in magnitude. This structure naturally suggests a sparsity correction. SCORE \citep{jinFastCommunityDetection2015} uses entry-wise ratios of eigenvectors to remove node-specific degree effects and has proved to be an effective tool for community detection under node degree heterogeneity. Motivated by this idea, we extend the ratio principle to multi-layer networks. In contrast to existing SCORE-type analyses for node heterogeneity, such as \cite{keSpecialInvitedPaper2023} and \cite{jinMixedMembershipEstimation2024}, our analysis establishes layer-clustering consistency under layer-wise sparsity heterogeneity and explicitly quantifies the effect of the largest and smallest layer sparsity levels.

In this paper, we propose the Heterogeneity-adjusted Iterative Network Tensor Spectral clustering (HINTS) method. HINTS first estimates the low-rank node and layer embeddings through a regularized tensor decomposition. A sparsity correction is then applied to the layer embedding before $K$-means clustering. The correction removes the layer-specific scale without estimating the individual sparsity parameters. Once the layer groups are recovered, HINTS performs community detection within each group and combines the local memberships to obtain a global community partition. We further establish theoretical guarantees for tensor decomposition, layer clustering, and community detection. The results keep track of the largest and smallest sparsity levels and characterize the range of their ratio under which consistent layer clustering can be achieved.

The main contributions of this paper are three-fold.

(1) We propose SH-MMSBM for multi-layer networks with both layer dependence and sparsity heterogeneity. It allows different layer groups to have different community structures, while layers within the same group may have different sparsity levels. This extends the existing mixture multi-layer models by accommodating within-group variation in network sparsity.

(2) We propose HINTS for layer clustering and community detection under SH-MMSBM. The oracle tensor representation shows that layer-specific sparsity acts as a multiplicative scale on the layer embedding. A sparsity correction is applied to the layer embedding before clustering, which removes the layer-specific scale effect without estimating the individual sparsity parameters. Based on the recovered layer groups, HINTS further identifies local community structures and integrates them into a global partition.

(3) The theoretical guarantees for tensor decomposition, layer clustering, and community detection explicitly track the effects of the sparsity heterogeneity. In particular, the ratio between the largest and smallest sparsity levels is allowed to diverge with the network dimensions, provided that the accumulated network signal is sufficiently strong.

The remainder of this paper is organized as follows. Section 2 introduces the proposed SH-MMSBM and presents the HINTS methodology. Section 3 establishes the theoretical results. Section 4 presents comparative numerical studies. Section 5 applies HINTS to the BACI international trade data. Section 6 concludes the paper. The Supplementary Materials contain proofs of the theoretical results, additional results for the BACI application, and an application to single-cell Hi-C data.

\textbf{Notation.} For a matrix $\bX$, we denote its $(i,j)$‑th entry, $i$‑th column, and $j$‑th row by $\bX_{ij}$, $\bX_{i:}$, and $\bX_{:j}$, respectively. The $k$‑th largest singular value of $\bX$ is $\sigma_k(\bX)$. 
For a tensor $\cX\in\bbR^{n_1\times n_2\times n_3}$, we write its $i_1$‑th horizontal, $i_2$‑th lateral, and $i_3$‑th frontal slices as $\cX_{i_1,:,:}\in\bbR^{n_2\times n_3}$, $\cX_{:,i_2,:}\in\bbR^{n_1\times n_3}$, and $\cX_{:,:,i_3}\in\bbR^{n_1\times n_2}$, respectively. The fibers $\cX_{:,i_2,i_3}\in\bbR^{n_1}$, $\cX_{i_1,:,i_3}\in\bbR^{n_2}$ and $\cX_{i_1,i_2,:}\in\bbR^{n_3}$ correspond to the mode‑1, mode‑2 and mode‑3 fibers. The $k$‑mode unfolding of $\cX$ is denoted by $\bbM_k(\cX)\in\bbR^{n_k\times\Pi_{j\neq k}n_j}$. 
The operator norm and Frobenius norm are $\|\cdot\|$ and $\|\cdot\|_F$, respectively. For $\bX\in\bbR^{n_1\times n_2}$, we define $\|\bX\|_{2\rightarrow\infty}=\max_{i\in[n_1]}\|\bX_{i,:}\|$. 
For two positive sequences $\{a_n\}$ and $\{b_n\}$, we write $a_n\asymp b_n$ if there exist constants $0<C_1<C_2$ such that $C_1a_n\le b_n\le C_2a_n$; $a_n\lesssim b_n$ ($a_n\gtrsim b_n$) if $a_n\le Cb_n$ ($a_n\ge Cb_n$) for some $C>0$; and $a_n\gg b_n$ if $b_n/a_n=o(1)$.

\section{Methodology}\label{sec:2}
\subsection{SH-MMSBM and Low-Rank Representation}

Suppose that the $L$ layers of a multi-layer network can be partitioned into $M$ groups. Layers belonging to group $m$ share a node-membership matrix $\bZ_m\in\{0,1\}^{N\times K_m}$ and a baseline connection-probability matrix $\bB_0^{(m)}\in[0,1]^{K_m\times K_m}$, where $K_m$ is the number of communities in group $m$. Each row of $\bZ_m$ contains exactly one non-zero entry, and
$
[\bZ_m]_{ik}=\bbI\bigl(\text{node }i\text{ belongs to community }k\text{ in group }m\bigr),
$
where $\bbI(\cdot)$ represents an indicator function. The rows of $\bZ_m$ are independently drawn from a multinomial distribution with probabilities $\bpi^{(m)}=(\pi_1^{(m)},\ldots,\pi_{K_m}^{(m)})^\top$, where $\sum_{k=1}^{K_m}\pi_k^{(m)}=1$.
For $l\in[L]$, let $g_l\in[M]$ denote the group label of layer $l$, with $\bbP(g_l=m)=\pi_m^\prime$. Write $\bpi^\prime=(\pi_1^\prime,\ldots,\pi_M^\prime)^\top$ with $\sum_{m=1}^M\pi_m^\prime=1$. The layer-membership matrix $\bG\in\{0,1\}^{L\times M}$ is defined by $\bG_{lm}=\bbI(g_l=m)$, and $\bg=(g_1,\ldots,g_L)^\top$ collects the layer labels.

To allow network sparsity to vary within a layer group, we introduce layer-specific sparsity parameters $\brho=(\rho_1,\ldots,\rho_L)^\top$, where $\rho_l\in(0,1)$. Conditional on $g_l=m$, layer $l$ follows
$$
\mathcal{G}_l\mid g_l=m\sim\operatorname{SBM}(\bZ_m,\rho_l\bB_0^{(m)}).
$$
For identifiability, we set $\max_{k_1,k_2\in[K_m]}[\bB_0^{(m)}]_{k_1k_2}=1$, for $m\in[M]$. Thus, \(\rho_l\) captures layer-specific variation in edge density.
Equivalently, conditional on the layer and node memberships,
$$
\cA_{i_1,i_2,l}\sim\operatorname{Bernoulli}\left(\rho_l[\bB_0^{(m)}]_{k_1k_2}\right),
$$
independently for $i_1\leq i_2$, where $g_l=m$, $c_{i_1}^{(m)}=k_1$, $c_{i_2}^{(m)}=k_2$, and $c_i^{(m)}$ denotes the community label of node $i$ in group $m$. We refer to this model as the Sparsity Heterogeneous Mixture Multi-layer Stochastic Block Model (SH-MMSBM).

Next, we concatenate the local community structures across layer groups to a global representation. Let $\mr\bZ=[\bZ_1,\ldots,\bZ_M]\in\{0,1\}^{N\times\dK}$, where $\dK=\sum_{m=1}^M K_m.$
The global number of communities $\rK$ is the number of distinct rows of $\mr\bZ$, and the corresponding global community label is $\bc=(c_1,\ldots,c_N)^\top\in[\rK]^N$. Hence,
$\max_{m\in[M]}K_m\leq\rK\leq\prod_{m\in[M]}K_m$. Let $r=\operatorname{rank}(\mr\bZ)$. In general, $r$ and $\rK$ need not be equal, and $\max_{m\in[M]}K_m\leq r\leq\min\{\rK,\dK-(M-1)\}$.
Take the compact singular‑value decomposition (SVD)
$
\mr\bZ=\bU_Z\bLambda_Z\bV_Z^\top,
$
where $\bU_Z\in\bbR^{N\times r}$ and $\bV_Z\in\bbR^{\dK\times r}$ have orthonormal columns and $\bLambda_Z\in\bbR^{r\times r}$ contains the non-zero singular values.
Define $\cB_0\in[0,1]^{\dK\times\dK\times M}$ so that the $m$-th frontal slice contains $\bB_0^{(m)}$ in the $m$-th diagonal block and zeros elsewhere. Let 
$\bR=\diag(\rho_1,\ldots,\rho_L)$, $\bGamma_3=\diag(\sqrt{\gamma_1},\ldots,\sqrt{\gamma_M})$, where $\gamma_m=\sum_{l=1}^L\rho_l^2\bG_{lm}$, for $m\in[M]$.

The oracle adjacency tensor $\cP=\bbE(\cA)$ can be written as
\begin{equation*}
\begin{aligned}
\cP=&(\cB_0\times_3\bGamma_3)\times_1\mr\bZ\times_2\mr\bZ\times_3\bR\bG\bGamma_3^{-1}\\
=&(\cC\times_1\bU_B\times_2\bU_B\times_3\bW_B)\times_1\bU_Z\times_2\bU_Z\times_3\bR\bG\bGamma_3^{-1}\\
=&\cC\times_1\bU_Z\bU_B\times_2\bU_Z\bU_B\times_3\bR\bG\bGamma_3^{-1}\bW_B\\
\triangleq&\cC\times_1\bU\times_2\bU\times_3\bW,
\end{aligned}
\end{equation*}
where $\cC\times_1\bU_B\times_2\bU_B\times_3\bW_B$ is the Tucker decomposition of $\cB_0\times_1\bLambda_Z\bV_Z^\top\times_2\bLambda_Z\bV_Z^\top\times_3\bGamma_3$. Here, $\cC\in\bbR^{r\times r\times M}$ is the core tensor, $\bU_B\in\bbR^{r\times r}$ and $\bW_B\in\bbR^{M\times M}$ are orthogonal matrices. $\bU=\bU_Z\bU_B \in \mathbb{R}^{N \times r}$ and $\bW=\bR\bG\bGamma_3^{-1}\bW_B\in \mathbb{R}^{L \times M}$ are the oracle node and layer embeddings, respectively.

The form of $\bW$ is particularly important for layer clustering. For layer $l$ with $g_l=m$,
\begin{equation}\label{eq:W_layer_geometry}
\bW_{l,:}=\rho_l\gamma_m^{-1/2}[\bW_B]_{m,:}=\rho_l[\bGamma_3^{-1}\bW_B]_{m,:}.
\end{equation}
Thus, layers in the same group share the same direction in the oracle layer embedding, while their magnitudes vary with $\rho_l$. When the sparsity levels are heterogeneous, rows from one group need not concentrate around a single Euclidean center. Directly applying $K$-means to $\bW$ can therefore be affected by the variation in network sparsity. Equation \eqref{eq:W_layer_geometry} also suggests a direct correction. Motivated by the ratio idea of SCORE \citep{jinFastCommunityDetection2015}, define the oracle sparsity-corrected embedding $\bS=\diag(\bW_{:,1})^{-1}\bW_{:,2:M}$. Both $\rho_l$ and the group-wise scaling factor $\gamma_{g_l}^{-1/2}$ cancel in the ratios. Thus,
\begin{equation*}
\bS_{l,:}=\left[\frac{[\bW_B]_{g_l,2}}{[\bW_B]_{g_l,1}},\ldots,\frac{[\bW_B]_{g_l,M}}{[\bW_B]_{g_l,1}}\right], \quad l\in[L].
\end{equation*}
The sparsity correction therefore removes the layer-specific scale without estimating $\rho_l$, and $\bS_{l,:}$ depends on layer $l$ only through its group label $g_l$.

In practice, given an estimate $\wh\bW$ of the layer embedding, we use the empirical counterpart
\begin{equation}\label{eq:empirical_SCORE}
\wt\bS=\diag(\wh\bW_{:,1})^{-1}\wh\bW_{:,2:M}.
\end{equation}
To control unstable ratios, each row is truncated as
\begin{equation}\label{eq:empirical_SCORE_truncation}
\wh\bS_{l,:}=\wt\bS_{l,:}\min\left\{1,\frac{\sqrt{M\log L}}{\|\wt\bS_{l,:}\|}\right\},\qquad l\in[L].
\end{equation}
As shown in the proof of Theorem \ref{thm:consistency_layer}, the oracle rows satisfy $\|\bS_{l,:}\|\lesssim\sqrt{M}$. The threshold $\sqrt{M\log L}$ leaves the oracle scale unchanged while controlling noise amplification caused by the ratio operation.

\begin{figure}[h]
    \centering
    \includegraphics[width=\linewidth]{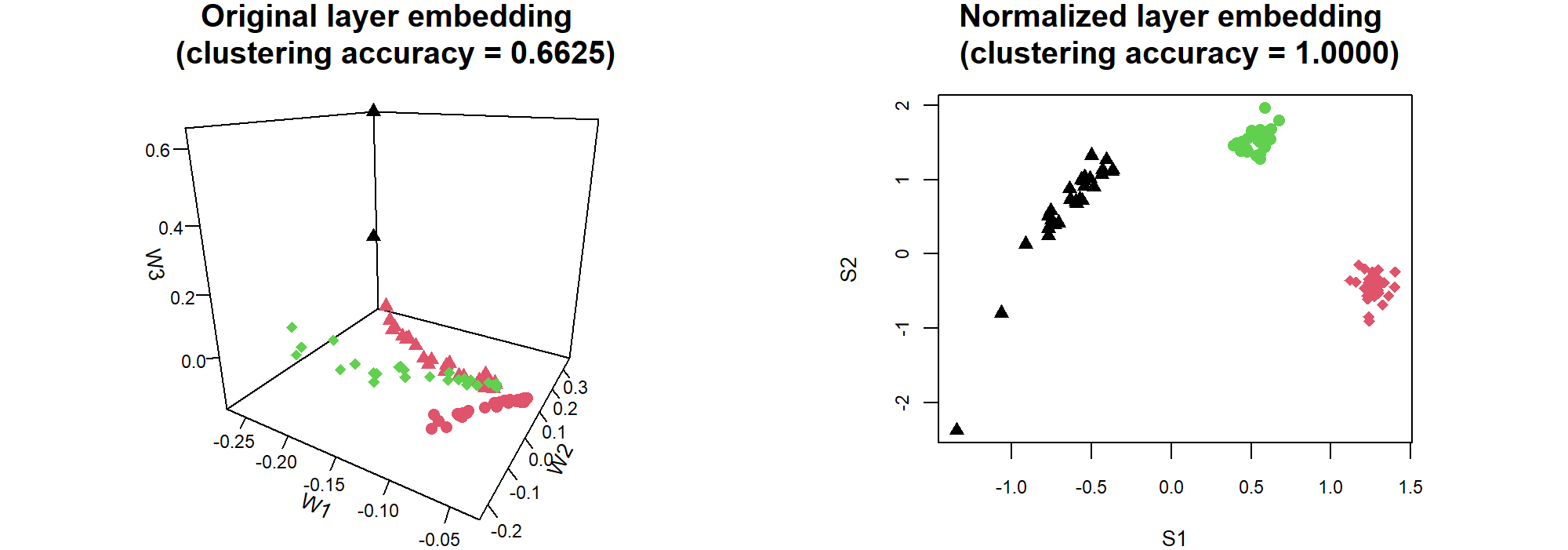}
    \caption{
    Network generated under SH-MMSBM as detailed in Section \ref{sec:4} through connection model (B2) with parameters $(L,N,M,\lambda,\alpha_1,\alpha_2)=(80,800,3,0.6,0.02,0.08)$ and $K_1=K_2=K_3=3$. Points represent the estimated layer embeddings: (left) rows of $\wh\bW$; (right) rows of the sparsity-corrected embedding $\wh\bS$. Points are shaped by true membership and colored by clustering result.}
    \label{fig:demo2_SCORE_normalization}
\end{figure}

A toy example shown in Figure~\ref{fig:demo2_SCORE_normalization} illustrates that the raw embeddings from the same group spread mainly along a common direction, whereas the corrected embeddings concentrate around their group centers. In this example, $K$-means applied to $\wh\bW$ gives an accuracy of $0.6625$, while $K$-means applied to $\wh\bS$ perfectly recovers the layer groups.

\subsection{Sparsity-Corrected Tensor Clustering}

We now describe HINTS for estimating the layer memberships and the local and global community structures from the observed tensor $\cA$. The procedure first estimates the low-rank tensor embeddings, then performs sparsity-corrected layer clustering, and finally recovers the community structures based on the estimated layer groups. 

We estimate $\bU$ and $\bW$ by a regularized Tucker iteration based on HOOI \citep{delathauwerBestRank1Rank2000}. A suitable initialization is important for the iterative refinement. Standard HOSVD \citep{delathauwerMultilinearSingularValue2000} initializes the factors using singular vectors of tensor unfoldings. For Bernoulli network data, the noise variance varies across tensor entries, which induces heterogeneous diagonal noise in the Gram matrices of the unfoldings. To obtain a warm start, we therefore initialize the mode-1 and mode-3 factors using HeteroPCA \citep{zhangHeteroskedasticPCAAlgorithm2022}. The complete initialization procedure is given in Appendix A.3.1 in the Supplementary Materials. 

The initial estimates are refined by power iteration with incoherence regularization. The incoherence condition is standard in low-rank tensor estimation \citep{keCommunityDetectionHypergraph2020,jingCommunityDetectionMixture2021,zhangEfficientEstimationLongitudinal2025} and prevents the estimated factors from concentrating on a small number of rows.

\begin{definition}[Incoherence parameters] 
Let $\delta_U=\|\bU\|_{2\rightarrow\infty}$, $\delta_W=\|\bW\|_{2\rightarrow\infty}$ be the incoherence parameters of the oracle node and layer embeddings. \end{definition} 

Proposition~\ref{prop:incoherence_condition} gives upper bounds for $\delta_U$ and $\delta_W$. In practice, we use $$ \wh\delta_U=\frac{10\sqrt{\rK}\max_{i\in[N]}d_{\operatorname{node}}(i)}{\sqrt{\sum_{i=1}^N d_{\operatorname{node}}^2(i)}},\qquad \wh\delta_W=\frac{10\rK\max_{l\in[L]}d_{\operatorname{layer}}(l)}{\sqrt{M\sum_{l=1}^L d_{\operatorname{layer}}^2(l)}}, $$ where $$ d_{\operatorname{node}}(i)=\sum_{j=1}^N\sum_{l=1}^L\cA_{ijl},\qquad d_{\operatorname{layer}}(l)=\sum_{i=1}^N\sum_{j=1}^N\cA_{ijl}. $$

For a matrix $\bX$ and threshold $\delta$, the incoherence regularization is achieved by
$$ \bar\bX_{i,:}=\bX_{i,:}\min\left\{1,\frac{\delta}{\|\bX_{i,:}\|}\right\}. $$
The operator $\Reg(\bX,\delta)$ returns the left singular matrix of $\bar\bX$. At iteration $t$, the node embedding is updated by \begin{equation}\label{eq:HINTS_U_update} 
\begin{aligned}
    \wt\bU^{(t)}=&\text{top-$r$ left singular vectors of }\bbM_1\left(\cA\times_2\wh\bU^{(t-1)\top}\times_3\wh\bW^{(t-1)\top}\right),\\
\wh\bU^{(t)}=&\Reg(\wt\bU^{(t)},\wh\delta_U),
\end{aligned}
\end{equation}
and the layer embedding is updated by \begin{equation}\label{eq:HINTS_W_update}
\begin{aligned}
\wt\bW^{(t)}=&\text{top-$M$ left singular vectors of }\bbM_3\left(\cA\times_1\wh\bU^{(t-1)\top}\times_2\wh\bU^{(t-1)\top}\right),\\
\wh\bW^{(t)}=&\Reg(\wt\bW^{(t)},\wh\delta_W).
\end{aligned}
\end{equation}
The iteration continues until convergence or until $T_{\max}$ iterations are reached, producing the final estimates $\wh\bU$ and $\wh\bW$.

With the tensor embeddings obtained, we next recover the layer memberships. The oracle analysis in \eqref{eq:W_layer_geometry} shows that the magnitude of the layer embedding is affected by the layer-specific sparsity parameter. To obtain the sparsity-corrected embedding depends only on the layer group, we construct $\wh\bS$ according to \eqref{eq:empirical_SCORE}--\eqref{eq:empirical_SCORE_truncation} and apply $K$-means to its rows with $M$ clusters. The resulting labels are denoted by $\wh\bg$. In contrast to direct clustering of $\wh\bW$, this step removes the layer-specific scale before comparing different layers. 

After the layer groups are recovered, we estimate the community structure within each group. For $m\in[M]$, let $\wh\cL_m=\{l\in[L]:\wh g_l=m\}$, and define the group-wise mean adjacency matrix 
\begin{equation}\label{eq:Ahat_group_mean} 
\wh\bA_m=\frac{1}{\#(\wh\cL_m)}\sum_{l\in\wh\cL_m}\cA_{:,:,l}. \end{equation}
Under the true layer grouping, $$ \bbE\left(\frac{1}{L_m}\sum_{l:g_l=m}\cA_{:,:,l}\right)=\bar\rho_m\bZ_m\bB_0^{(m)}\bZ_m^\top, $$ where $L_m=\#\{l:g_l=m\}$ and $\bar\rho_m=L_m^{-1}\sum_{l:g_l=m}\rho_l$. Hence, averaging the layers within a group preserves its community structure while pooling information across the corresponding layers. We apply spectral clustering to $\wh\bA_m$ with $K_m$ communities to obtain the local membership estimate $\wh\bZ_m$ and the local label $\wh\bc^{(m)}$.

\begin{figure}[ht] 
\centering 
\includegraphics[width=\linewidth]{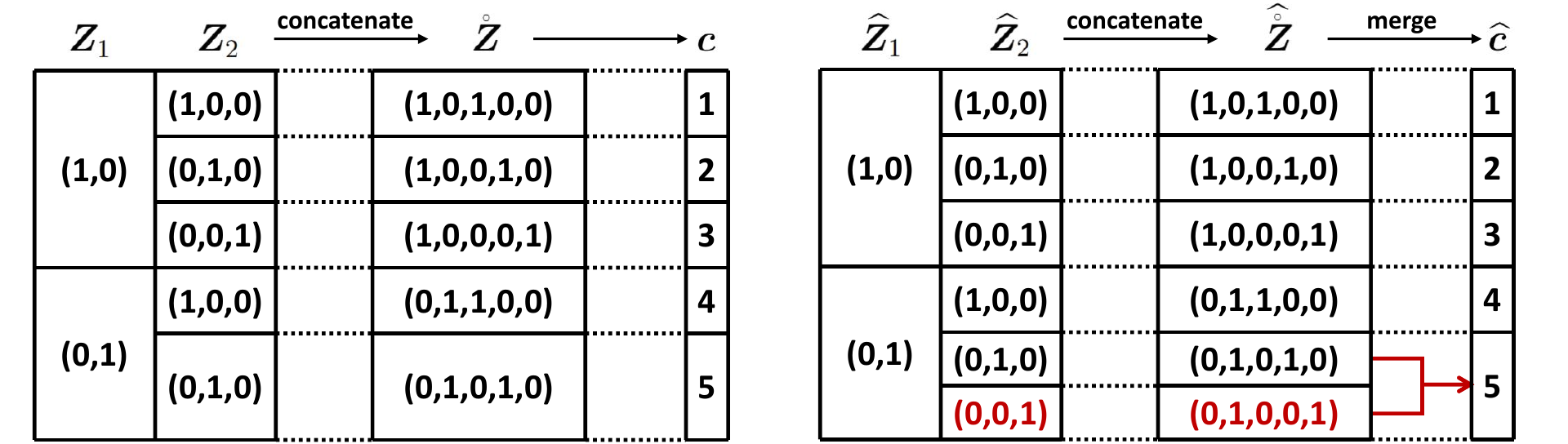} 
\caption{Illustration of local and global community detection.}
\label{fig:demo3} 
\end{figure}

The local memberships are then combined to recover the global community structure. Define
\(
\wh{\mr\bZ}=[\wh\bZ_1,\ldots,\wh\bZ_M].
\)
Nodes sharing the same row of $\wh{\mr\bZ}$ form an estimated global pattern group. When more than $\rK$ distinct patterns are present, we retain the $\rK$ largest pattern groups and merge the remaining groups into them using centroid distances in $\wh\bU$. Figure~\ref{fig:demo3} illustrates this construction.

The complete procedure is summarized in Algorithm~\ref{al:HINTS}. 

\begin{algorithm}[H] 
\baselineskip=14pt 
\caption{HINTS}\label{al:HINTS} 
\begin{algorithmic}[1] 
\Require Observed adjacency tensor $\cA\in\{0,1\}^{N\times N\times L}$, Tucker rank $(r,r,M)$, number of layer groups $M$, numbers of local communities $\{K_1,\ldots,K_M\}$, number of global communities $\rK$, incoherence thresholds $\delta_U,\delta_W$, and maximum iterations $T_{\max},T'_{\max}$. 
\Ensure Estimated layer labels $\wh\bg$, local community labels $\{\wh\bc^{(1)},\ldots,\wh\bc^{(M)}\}$, and global community labels $\wh\bc$. 
\State Initialize $\wh\bU^{(0)}$ and $\wh\bW^{(0)}$ with HeteroPCA. 
\State $t\gets0$. 
\Repeat 
\State $t\gets t+1$. 
\State Update $\wh\bU^{(t)}$ and $\wh\bW^{(t)}$ according to \eqref{eq:HINTS_U_update} and \eqref{eq:HINTS_W_update}. 
\Until{convergence or $t=T_{\max}$.} 
\State Set $\wh\bU\gets\wh\bU^{(t)}$ and $\wh\bW\gets\wh\bW^{(t)}$. 
\State Construct $\wh\bS$ according to \eqref{eq:empirical_SCORE}--\eqref{eq:empirical_SCORE_truncation} and apply $K$-means to obtain $\wh\bg$. 
\For{$m=1$ to $M$} 
\State Construct $\wh\bA_m$ according to \eqref{eq:Ahat_group_mean} and apply spectral clustering with $K_m$ communities to obtain $\wh\bZ_m$ and $\wh\bc^{(m)}$. 
\EndFor 
\State Set $\wh{\mr\bZ}\gets[\wh\bZ_1,\ldots,\wh\bZ_M]$. 
\If{the number of distinct rows of $\wh{\mr\bZ}$ exceeds $\rK$} 
\State Retain the $\rK$ largest distinct-row groups of $\wh{\mr\bZ}$ and merge the remaining groups into them using centroid distances in $\wh\bU$; obtain $\wh\bc$.
\Else
\State Obtain $\wh\bc$ from the distinct row patterns of $\wh{\mr\bZ}$. 
\EndIf
\end{algorithmic}
\end{algorithm} 

The structural quantities in Algorithm~\ref{al:HINTS} can be selected in a data-driven manner. In practice, $M$ and $r$ can be estimated from the effective ranks of the mode-3 and mode-1 unfoldings using singular-value scree or eigen-ratio rules. The local community numbers $K_m$ can be selected by eigen-ratio, ECV in \citep{liNetworkCrossvalidationEdge2020} or StGoF in \citep{jinOptimalEstimationNumber2022}. The global number $\rK$ can be obtained from stable combinations of the local membership patterns, or supplied from prior information when available, or estimated with the same method as $r$.

\section{Theoretical Results}
\subsection{Assumptions and Propositions}

\begin{assumption}\label{ass:HINTS:1}
For SH-MMSBM, we assume that
\begin{itemize}
    \item[\upshape{(1)}] Define $\mr\cB_0=\cB_0\times_{1}\bV_Z^\top\times_{2}\bV_Z^\top\in\bbR^{r\times r\times M}$. There exists a constant $b>0$ such that $\min_{j\in[3]}\{\sigma_{\min}(\bbM_j(\mr\cB_0))\}\geq b$, where $\sigma_{\min}(\cdot)$ denotes the minimum nonzero singular value;
    \item[\upshape{(2)}] There exists a constant $\kappa\geq1$ such that $\sigma_1(\mr\bZ)\leq\kappa\sigma_r(\mr\bZ)$.
\end{itemize}
\end{assumption}

Assumption~\ref{ass:HINTS:1} imposes standard non-degeneracy conditions on the oracle structure. Condition (1) requires the projected connectivity tensor to contain non-vanishing information along each mode, which is needed for identifying the corresponding low-rank tensor factors. Condition (2) requires $\mr\bZ$ to be well-conditioned and prevents the global membership structure from becoming nearly rank deficient. This condition is related to the number and sizes of the global communities and is commonly needed to maintain sufficient separation in spectral methods.

\begin{assumption}\label{ass:HINTS:2}
Regarding the layer groups, we assume that
\begin{itemize}
    \item[\upshape{(1)}] the group-wise sparsity satisfies $\gamma_{\min}\asymp\gamma_{\max}$, where $\gamma_{\max}=\max_{m\in[M]}\gamma_m$ and $\gamma_{\min}=\min_{m\in[M]}\gamma_m$;
    \item[\upshape{(2)}] the group size satisfies $L_{\max}\asymp L_{\min}\asymp L/M$, where $L_{\min}=\min_{m\in[M]}\left(\sum_{l=1}^L\bG_{lm}\right)$ and $L_{\max}=\max_{m\in[M]}\left(\sum_{l=1}^L\bG_{lm}\right)$.
\end{itemize}
\end{assumption}

Assumption~\ref{ass:HINTS:2} requires balance only at the layer-group level. Condition (1) controls the aggregate squared sparsity $\gamma_m=\sum_{l:g_l=m}\rho_l^2$ across groups. Hence, considerable sparsity heterogeneity is still allowed across layers. Condition (2) rules out asymptotically vanishing layer groups and ensures that each group contains a non-negligible proportion of the layers. These conditions maintain comparable amounts of information across layer groups while leaving the within-group sparsity variation largely unrestricted.

\begin{proposition}\label{prop:signal_strenth}
Under Assumption~\ref{ass:HINTS:1}, the signal strength of the core tensor $\cC$ satisfies
$$
\min_{j\in[3]}\{\sigma_{\min}(\bbM_j(\cC))\}\geq\kappa^{-2}r^{-1}bMN\gamma_{\min}^{1/2}.
$$
\end{proposition}

Proposition~\ref{prop:signal_strenth} gives a lower bound for the singular values of the oracle core tensor that determine the difficulty of estimating its low-rank factors. The bound increases with the network size $N$ and with the cumulative sparsity level represented by $\gamma_{\min}$. In particular, when $\rho_{\max}\asymp\rho_{\min}\asymp\rho$ and the layer groups are balanced, the lower bound reduces to $C\kappa^{-2}r^{-1}bM^{1/2}\rho L^{1/2}N$ for some constant $C>0$. This agrees with the usual scaling of the oracle tensor strength in the homogeneous setting.

\begin{proposition}\label{prop:incoherence_condition}
Under Assumption~\ref{ass:HINTS:1}, the rows of $\bU$ and $\bW$ satisfy
$$
\|\bU_{i,:}\|\leq\kappa r^{1/2}N^{-1/2}\quad\text{and}\quad\|\bW_{l,:}\|\leq\kappa^2rb^{-1}M^{-1}\gamma_{\min}^{-1/2}\rho_l.
$$
Furthermore,
$$
\delta_U\leq\kappa r^{1/2}N^{-1/2}\quad\text{and}\quad\delta_W\leq\kappa^2rb^{-1}M^{-1}\gamma_{\min}^{-1/2}\rho_{\max}.
$$
\end{proposition}

Proposition~\ref{prop:incoherence_condition} provides a direct basis for choosing the incoherence thresholds in Algorithm~\ref{al:HINTS}. In particular, the row norm of $\bW$ varies with the layer sparsity $\rho_l$, which is consistent with the oracle geometry described in Section~\ref{sec:2}. The resulting bounds are also used in the subsequent analysis, which yield sharper control of the Bernoulli noise through incoherent tensor norm defined in \cite{yuanIncoherentTensorNorms2017}.

\subsection{Main Theorems}
We first study the convergence in estimation of the low-rank tensor embeddings, and then provide consistency theorems for clustering tasks. 

Let $\bbO_k$ denote the set of $k \times k$ orthonormal matrices. For matrices $\bX, \bX' \in \bbR^{n_1 \times r_1}$, we define the distance
$$
\d(\bX', \bX) = \inf_{\bO \in \bbO_{r_1}} \|\bX' - \bX \bO\|.
$$
Let $\d_{\max}^{(t)} = \max\left\{\d(\wh\bU^{(t)},\bU),\d(\wh\bW^{(t)},\bW)\right\}$ be the estimation error at iteration $t$. Let
$$
\nu_1=\left(\delta_Ur^{1/2}\rho_{\max}^{1/2}+\sqrt{\sum_{l=1}^L\rho_l(\delta_U^2\vee\|\bW_{l,:}\|^2)}\right)N^{1/2}\log^{1/2}(L\vee N)+\delta_U(\delta_U\vee\delta_W)\log(L\vee N).
$$

\begin{theorem}[Linear convergence of tensor decomposition]\label{thm:linear_convergence}
Suppose Assumption~\ref{ass:HINTS:1} holds and
$$
\begin{aligned}
\rho_{\max}LN&\gtrsim\log(L\vee N),\\
\rho_{\max}^{-1/2}\gamma_{\min}^{1/2}N(L\vee N)^{-1/2}L^{-1/2}L_{\min}^{1/2}&\gtrsim\kappa^4r^{5/2}\log(\kappa r)b^{-1}M^{-1/2}\log^2(L\vee N).
\end{aligned}
$$
Suppose further that the initialization satisfies $\d_{\max}^{(0)}\leq1/4$. Let
$$
t_{\max}=O\left(\log\left[\min_{j\in[3]}\{\sigma_{\min}(\bbM_j(\cC))\}/\nu_1\right]\vee1\right).
$$
Then, with probability $1-o((L+N)^{-3})$,
$$
\d_{\max}^{(t)}\leq\frac{1}{2}\d_{\max}^{(t-1)}+\frac{C\nu_1}{\min_{j\in[3]}\{\sigma_{\min}(\bbM_j(\cC))\}},\qquad 1\leq t\leq t_{\max},
$$
for some constant $C>0$, and
$$
\d_{\max}^{(t_{\max})}\lesssim\frac{\nu_1}{\min_{j\in[3]}\{\sigma_{\min}(\bbM_j(\cC))\}}.
$$
\end{theorem}

Theorem \ref{thm:linear_convergence} focuses on the iterative refinement and its final estimation error. At each iteration, the previous error is reduced by a factor of $1/2$ until it reaches the order $\nu_1/\min_{j\in[3]}\{\sigma_{\min}(\bbM_j(\cC))\}$. The conditions guarantee adequate signal-to-noise ratio to control the incoherent tensor operator norm of $\mathcal{A}-\mathcal{P}$. Although the theorem is stated conditional on $\d_{\max}^{(0)}\leq1/4$, Theorem A.1 in the Supplementary Materials guarantees this warm-initialization under suitable conditions. After a logarithmic number of iterations, the initialization error becomes negligible and the final rate is determined by the network sparsity, sparsity heterogeneity, and layer-group structure. Under these conditions, the embedding error satisfies $\mathrm{d}_{\max}^{(t_{\max})} \overset{P}{\rightarrow} 0$ as $L,N \to \infty$, confirming the consistency of the proposed estimator.

\begin{corollary}\label{cor:linear_covergence}
Suppose Assumptions~\ref{ass:HINTS:1}-\ref{ass:HINTS:2} and the conditions in Theorem \ref{thm:linear_convergence} hold, $N\geq L$, and $b,\kappa,r$ and $M$ are of constant order. If
$$
\rho_{\max}LN\gtrsim\log^4N, \qquad \frac{\rho_{\min}}{\rho_{\max}^{1/2}}L^{1/2}N^{1/2}\gg\log^{1/2}N,
$$
after $t_{\max}=O(\log N)$ iterations,
\begin{equation}\label{eq:cor_dmax}
\d_{\max}^{(t_{\max})}\lesssim\frac{\rho_{\max}^{1/2}}{\rho_{\min}}\frac{\log^{1/2}N}{L^{1/2}N^{1/2}}
\end{equation}
with probability $1-o((L+N)^{-3})$.
\end{corollary}

Corollary~\ref{cor:linear_covergence} gives a more transparent view of the effect of sparsity heterogeneity. The factor $\rho_{\max}^{1/2}/\rho_{\min}$ in \eqref{eq:cor_dmax} shows that tensor estimation becomes more difficult when the sparsity levels are more heterogeneous, while increasing $L$ or $N$ reduces this effect. In the homogeneous case $\rho_{\max}\asymp\rho_{\min}\asymp\rho$, the conditions reduce to $\rho LN\gtrsim\log^4N$ and the convergence rate becomes
$$
\frac{\log^{1/2}N}{\rho^{1/2}L^{1/2}N^{1/2}}.
$$
This agrees with the corresponding tensor-decomposition rate in Corollary 1 of \cite{jingCommunityDetectionMixture2021}. 

Next, we establish the clustering consistency. For the true label vector $\balpha \in \bbR^n$ and its estimator $\wh \balpha$, we quantify clustering error by the Hamming distance
$$
\err(\balpha,\wh\balpha)=\min_{\Pi}\frac{1}{n}\sum_{i=1}^n \bbI[\balpha_i \neq \Pi(\wh \balpha_i)],
$$
where $\Pi(\cdot)$ denotes a permutation of the label set.

\begin{theorem}[Consistency of layer clustering]\label{thm:consistency_layer} 
Suppose Assumptions \ref{ass:HINTS:1}-\ref{ass:HINTS:2} and all conditions in Theorem \ref{thm:linear_convergence} hold. We further assume that 
\begin{equation}\label{eq:thm_lc_sparsity}
\rho_{\min}\rho_{\max}^{-3/2} \gamma_{\min}^{1/2} N (L\vee N)^{-1/2} L^{-1/2} L_{\min}^{1/2} \gtrsim \kappa^4 r^{5/2} \log(\kappa r) b^{-1} \log^{2}(L\vee N)\log^{1/2}L
\end{equation}
and
\begin{equation}\label{eq:thm_lc_sigma12}
  \sigma_1(\bbM_3(\cC))- \sigma_2(\bbM_3(\cC)) \gg \nu_1 M^{1/2} \log^{1/2} L.
\end{equation}
Then, with probability $1-o((L+N)^{-3})$, the misclustering rate of layers satisfies
$$
\err(\bg,\wh\bg) \lesssim \frac{\rho_{\max}^2\nu_1^2}{\rho_{\min}^2\min\{\sigma_{\min}^2(\bbM_3(\cC)),\{\sigma_1(\bbM_3(\cC))-\sigma_2(\bbM_3(\cC))\}^2\}}.
$$
\end{theorem}

Theorem \ref{thm:consistency_layer} establishes the consistency of layer clustering. To understand the sparsity requirement, consider $N\geq L$ and let $b,\kappa,r$ and $M$ be of constant order. Then, condition \eqref{eq:thm_lc_sparsity} is satisfied when 
$$
\rho_{\min}\rho_{\max}^{-1/2}L^{1/2}N^{1/2} \gg \log^2(L\vee N) \log^{1/2} L,
$$
which covers a wide range of sparsity heterogeneity. Under the same setting, condition \eqref{eq:thm_lc_sparsity} can be equivalently written as
$$
\frac{\rho_{\max}}{\rho_{\min}}
\ll
\frac{\rho_{\min}LN}{\log^4N\log L}.
$$
Therefore, the ratio $\rho_{\max}/\rho_{\min}$ is allowed to diverge as $L$ and $N$ increase, provided that the information accumulated from the layers and nodes grows sufficiently fast. This explicitly characterizes the trade-off between network sparsity and sparsity heterogeneity. Relative to a homogeneous network with sparsity level $\rho_{\min}$, the tensor estimation error is enlarged by a factor $(\rho_{\max}/\rho_{\min})^{1/2}$. Thus, greater heterogeneity requires more layers or larger networks to achieve the same estimation accuracy.

Condition \eqref{eq:thm_lc_sigma12} is specific to the SCORE normalization. It requires the leading mode-3 singular direction to be sufficiently separated from the remaining directions so that the denominator in the eigenvector ratios can be estimated uniformly over layers, where a similar requirement appears in Theorem 1 of \cite{keCommunityDetectionHypergraph2020}. When $M$ is fixed and the spectral gap is of the same order as $\sigma_{\min}(\bbM_3(\cC))$, this introduces only a logarithmic strengthening over the signal condition required for tensor decomposition in Corollary \ref{cor:linear_covergence}.

The layer misclustering rate vanishes as the network dimensions $L$ and $N$ grow, which demonstrates that HINTS consistently recovers layer memberships even under heterogeneous sparsity. When $\rho_{\max}\asymp\rho_{\min}\asymp\rho$, condition \eqref{eq:thm_lc_sparsity} reduces to
$$
\rho LN\gg\log^4N\log L.
$$

Define $\err_L = C\rho_{\max}^2\rho_{\min}^{-2}\nu_1^2/\min\{\sigma_{\min}^2(\bbM_3(\cC)),\{\sigma_1(\bbM_3(\cC))-\sigma_2(\bbM_3(\cC))\}^2\}$, for some $C>0$, such that $\err(\bg,\wh\bg)\leq\err_L$ with probability $1-o((L+N)^{-3})$, as guaranteed by Theorem \ref{thm:consistency_layer}. For each layer group $m\in[M]$, define $L_m=\sum_{l=1}^L\bG_{lm}$, $\bar\rho_m=\sum_{l\in\cL_m}\rho_l/L_m$ and $N_k^{(m)}=\sum_{i \in [N]}[Z_m]_{ik}$. Denote $N_{\min}^{(m)}=\min_{k \in [K_m]} N_k^{(m)}$, $N_{\max}^{(m)}=\max_{k \in [K_m]} N_k^{(m)}$ and $K_{\max}=\max_{m\in[M]}K_m$. We further define
$$
\nu_U=\delta_U\nu_1+\left\{r\log(L\vee N)\sum_{l=1}^L\rho_l\|\bW_{l,:}\|^2\right\}^{1/2}
+\delta_U\delta_W\log(L\vee N).
$$
Then, we provide the consistency theorem for community detection.

\begin{theorem}[Consistency of community detection]\label{thm:consistency_community detection}

Suppose Assumptions \ref{ass:HINTS:1}-\ref{ass:HINTS:2} and all conditions in Theorem \ref{thm:consistency_layer} hold. Suppose $M\err_L=o(1)$ and
$$
\frac{\sigma_1(\mr\bZ)\nu_U}{\min_{j\in[3]}\{\sigma_{\min}(\bbM_j(\cC))\}}=o(1).
$$
For each $m\in[M]$, suppose
\[
\sigma_{\min}^2(\bB_0^{(m)})\bar\rho_mL_mN\gg\log(L\vee N).
\]
Then, with probability $1-o((L+N)^{-3})$, the misclustering rate of local community detection satisfies
$$
\begin{aligned}
\err(\bc^{(m)},\wh\bc^{(m)})
\lesssim&\frac{K_mN_{\max}^{(m)}N^2}{N_{\min}^{(m)3}\sigma_{\min}^2(\bB_0^{(m)})}
\left[
M^2\err_L^2\left\{\frac{\rho_{\max}}{\bar\rho_m}+\frac{\log(L\vee N)}{\bar\rho_mN}\right\}^2
+\frac{\log(L\vee N)}{\bar\rho_mL_mN}
\right].
\end{aligned}
$$
The misclustering rate of global community detection satisfies
$$
\begin{aligned}
\err(\bc,\wh\bc)\lesssim & K_{\max}N^2\max_{m\in[M]}\left(\frac{N_{\max}^{(m)}} {N_{\min}^{(m)3}\sigma_{\min}^2(\bB_0^{(m)})}\right.\\
&\left.\times\left[M^3\err_L^2\left\{\frac{\rho_{\max}}{\bar\rho_m}+\frac{\log(L\vee N)}{\bar\rho_mN}\right\}^2+\frac{M^2\log(L\vee N)}{\bar\rho_mLN}
\right]\right),
\end{aligned}
$$
with probability $1-o((L+N)^{-3})$.
\end{theorem}

Theorem \ref{thm:consistency_community detection} gives consistency guarantees for both local and global community detection. By explicitly featuring $K_{m}$, $N_{\max}^{(m)}$ and $N_{\min}^{(m)}$, the result accommodates moderate imbalance in community sizes and even permits $K_m$ to diverge with the network size. The local error bound improves as the cumulative signal $\bar\rho_m L_m$ increases, as the number of communities $K_m$ decreases, and as the community sizes $N_{\max}^{(m)}$ and $N_{\min}^{(m)}$ become more balanced.

The global error rate depends on the accuracy of the local membership estimates. Consider the special case in which $N_{\min}^{(m)}\asymp N_{\max}^{(m)}\asymp N/\rK$, $r\asymp\rK \asymp 1$, $\rho_{\max}\asymp\rho_{\min}\asymp\rho$ and $\kappa,b,M,\sigma_{\min}^2(\bB_0^{(m)})\asymp1$. In this regime, the additional sparsity requirement for community detection reduces to $\rho LN\gg\log N$, and the global error rate is of order
$$
\err_L+\frac{\log N}{\rho LN}.
$$
The second term agrees with the corresponding rates in Theorem 5.2 of \cite{jingCommunityDetectionMixture2021} and the proof of Theorem 1 in \cite{xuCovariateassistedCommunityDetection2022}. Thus, conditional on sufficiently accurate layer clustering, HINTS retains the error rate of existing tensor-based community detection methods. The overall rate additionally contains $\err_L$, reflecting the propagation of layer-clustering error to community detection.

\section{Numerical Studies}\label{sec:4}

We conduct numerical experiments to evaluate HINTS for both layer clustering and global community detection. We compare the proposed HINTS with two tensor-decomposition based methods, HOOI and TWIST. For a fair comparison, all iterative methods use the same HOSVD+HeteroPCA initialization, and we also report the clustering result obtained directly from this initialization. The performance is measured by the layer clustering accuracy $1-\err(\bg,\wh\bg)$ and the global community detection accuracy $1-\err(\bc,\wh\bc)$, averaged over 100 independent repetitions.

Multi-layer networks are generated from SH-MMSBM. For a network with $N$ nodes and $L$ layers, the $M$ layer groups are balanced with $\pi_1^\prime=\cdots=\pi_M^\prime=1/M$. Within each group $m$, the $K_m$ communities are also balanced, satisfying $\pi_1^{(m)}=\cdots=\pi_{K_m}^{(m)}=1/K_m$. To generate heterogeneous network sparsity, we first draw $\eta_l\sim U(0,1)$ independently and then set
$$
\rho_l=\left\{
\begin{aligned}
&\alpha_1+(\alpha_2-\alpha_1)\eta_l,\qquad &&\text{with probability }0.25,\\
&\alpha_1+(\alpha_2-\alpha_1)\eta_l^2,\qquad &&\text{with probability }0.75,
\end{aligned}
\right.
$$
for $l\in[L]$, where $0<\alpha_1<\alpha_2$ controls the range of layer sparsity.

We consider two constructions for the baseline probability matrices $\{\bB_0^{(m)}\}_{m=1}^M$.
\begin{itemize}
    \item[(B1)] Planted partition model: for $m\in[M]$ and $k_1,k_2\in[K_m]$, let
    $$
    [\bB_0^{(m)}]_{k_1,k_2}=p\bbI(k_1=k_2)+\lambda p\bbI(k_1\neq k_2).
    $$
    \item[(B2)] Multiplicative structure model: for each $m\in[M]$, independently generate $\bb_k^{(m)}\sim U(p_1,p_2)$ for $k\in[K_m]$ and let
    $$
    [\bB_0^{(m)}]_{k_1,k_2}=\bb_{k_1}^{(m)}\bb_{k_2}^{(m)}\bbI(k_1=k_2)+\lambda\bb_{k_1}^{(m)}\bb_{k_2}^{(m)}\bbI(k_1\neq k_2).
    $$
\end{itemize}
Model (B1) has a common within-community probability, whereas (B2) introduces community-specific interaction intensities and therefore allows more variation in the baseline connectivity pattern. Throughout the experiments, we set $p=0.6$ for (B1) and $(p_1,p_2)=(0.3,0.9)$ for (B2).

\begin{figure}[H]
    \centering
    \includegraphics[width=\linewidth]{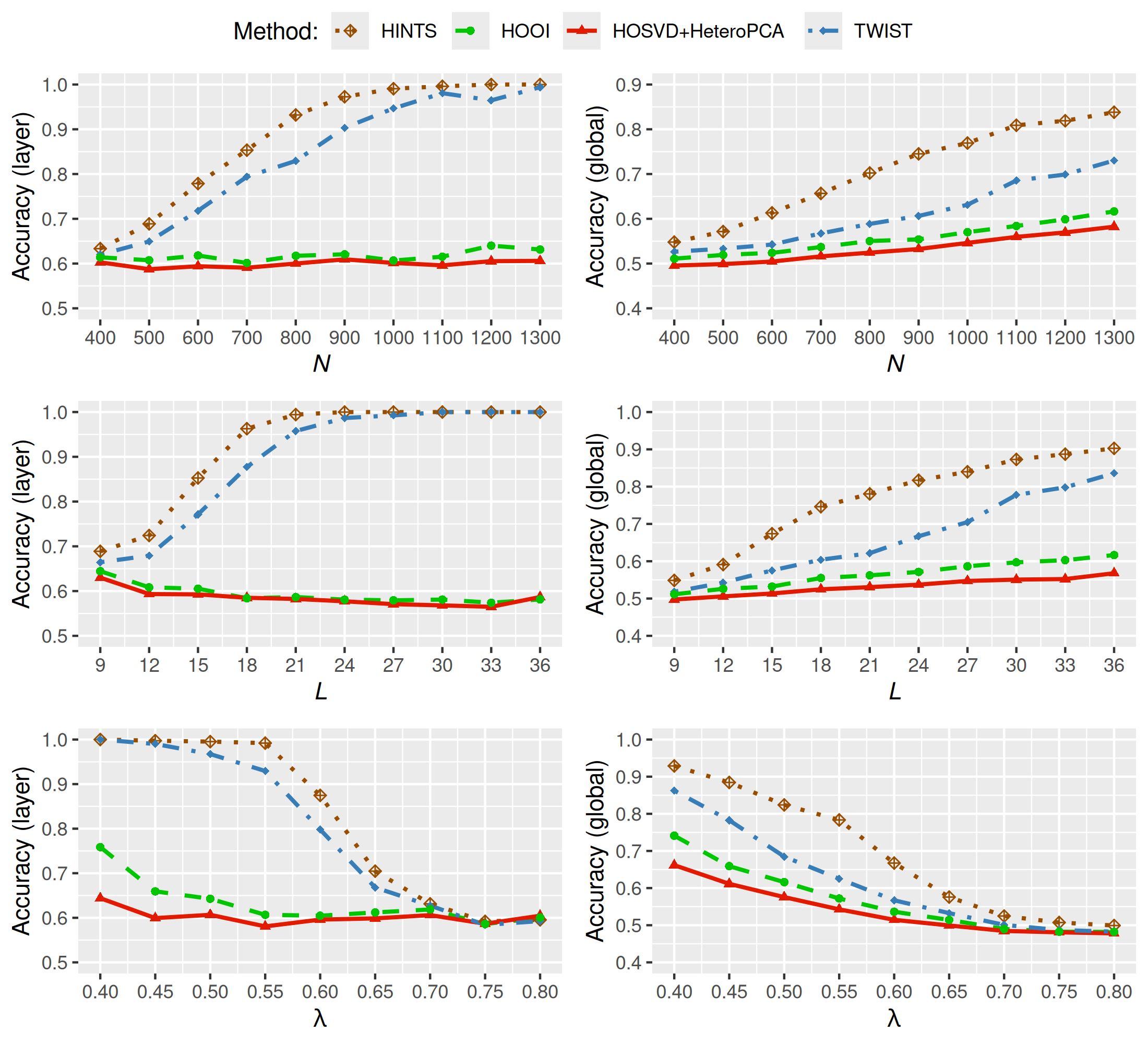}
    \caption{Layer clustering accuracy (left) and global community detection accuracy (right) in Simulation 1 under model (B1), as $N$, $L$, and $\lambda$ vary, respectively.}
    \label{fig:simul11}
\end{figure}

\begin{figure}[H]
    \centering
    \includegraphics[width=\linewidth]{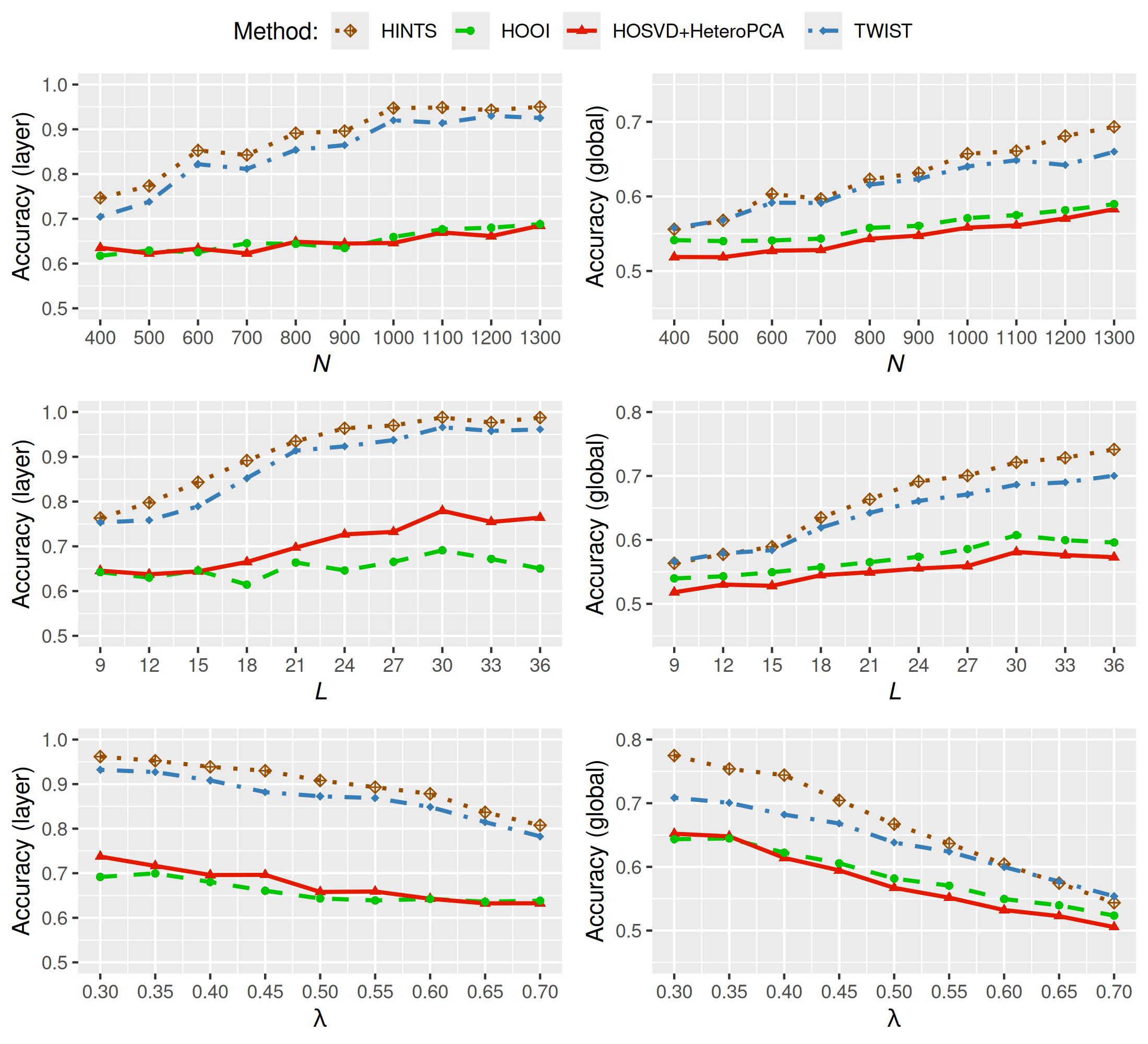}
    \caption{Layer clustering accuracy (left) and global community detection accuracy (right) in Simulation 1 under model (B2), as $N$, $L$, and $\lambda$ vary, respectively.}
    \label{fig:simul12}
\end{figure}

\textbf{Simulation 1.} We first examine how the performance changes with the network dimensions and the separation between communities. The baseline setting is $(L,N,M,\lambda,\alpha_1,\alpha_2)=(15,700,2,0.6,0.02,0.04)$ with $(K_1,K_2)=(2,3)$. We vary $N$, $L$, and $\lambda$ separately while fixing the remaining parameters at their baseline values. Figures~\ref{fig:simul11} and \ref{fig:simul12} report the results under models (B1) and (B2), respectively. In each figure, the left panels show layer clustering accuracy and the right panels show global community detection accuracy.

Several patterns are consistent across the two models. First, the accuracies of HINTS and TWIST improve steadily as $N$ or $L$ increases. For layer clustering, both methods approach perfect recovery when the network contains sufficiently many nodes or layers, while HOOI and HOSVD+HeteroPCA remain much less accurate. The improvement with $N$ and $L$ is also clear for global community detection. These results agree with the theoretical analysis, where information accumulated across nodes and layers improves both layer and community recovery.

Second, the performance deteriorates as $\lambda$ increases. A larger $\lambda$ makes between-community connections closer to within-community connections, so the community structures become less distinguishable. This affects not only community detection but also layer clustering, since the layer groups are defined through their underlying community structures. 

Compared with TWIST, the proposed HINTS further improves layer clustering in most nontrivial settings, and the improvement is also propagated to global community detection. The same pattern appears under both (B1) and the more heterogeneous baseline structure (B2). These comparisons indicate that correcting the layer embedding for sparsity variation is valuable.

\textbf{Simulation 2.} We focus directly on the effect of network sparsity and sparsity heterogeneity. The baseline setting is $(L,N,M,\lambda)=(30,1000,3,0.6)$ with $(K_1,K_2,K_3)=(2,2,3)$, and $\bB_0$ is generated from model (B2). We vary $(\alpha_1,\alpha_2)$ while keeping all other parameters fixed. The results are summarized in Table~\ref{tab:simul2}.

\begin{table}[htbp]
\footnotesize
\centering
\caption{Comparison results in Simulation 2.}
\label{tab:simul2}
\begin{tabularx}{\textwidth}{llccccc}
\hline
\multirow{2}{*}{Accuracy} & \multirow{2}{*}{Method} & \multicolumn{5}{c}{Sparsity parameters $(\alpha_1,\alpha_2)$} \\
\cline{3-7}
 & & (0.02,0.04) & (0.02,0.08) & (0.04,0.06) & (0.04,0.10) & (0.06,0.08) \\
\hline
\multirow{4}{*}{Layer}
&HINTS & 0.839 & 0.908 & 0.958 & 0.960 & 0.986 \\
&HOSVD+HeteroPCA & 0.604 & 0.542 & 0.680 & 0.578 & 0.710 \\
&HOOI & 0.578 & 0.621 & 0.729 & 0.742 & 0.779 \\
&TWIST & 0.822 & 0.835 & 0.952 & 0.938 & 0.976 \\
\hline
\multirow{4}{*}{Global}
&HINTS & 0.704 & 0.769 & 0.800 & 0.860 & 0.839 \\
&HOSVD+HeteroPCA & 0.625 & 0.662 & 0.674 & 0.700 & 0.699 \\
&HOOI & 0.640 & 0.667 & 0.681 & 0.699 & 0.714 \\
&TWIST & 0.683 & 0.705 & 0.726 & 0.744 & 0.748 \\
\hline
\end{tabularx}
\end{table}

The choices of $(\alpha_1,\alpha_2)$ vary both the overall sparsity level and the amount of heterogeneity. To separate these two effects, we first consider the settings $(0.02,0.04)$, $(0.04,0.06)$, and $(0.06,0.08)$, which have the same range $\alpha_2-\alpha_1=0.02$. As the network becomes denser, both layer and global accuracies generally increase. Among all the methods, HINTS achieves the highest accuracy in all the settings.

We next fix the average sparsity level and compare different amounts of heterogeneity. The settings $(0.02,0.08)$ and $(0.04,0.06)$ have the same midpoint of the sparsity range. Reducing the heterogeneity improves the layer accuracy of HINTS from $0.908$ to $0.958$ and the global accuracy from $0.769$ to $0.800$. The layer results show the same pattern for $(0.04,0.10)$ and $(0.06,0.08)$. The global accuracy in this pair is not monotonic, indicating that finite-sample community recovery also depends on the particular realized network structure. Overall, the results support the theoretical conclusion that greater sparsity heterogeneity makes recovery more difficult, while sufficient network sparsity can compensate for this effect.

HINTS achieves the highest layer clustering accuracy in all five settings and also gives the highest global accuracy. Its advantage over TWIST is especially clear in the more heterogeneous settings $(0.02,0.08)$ and $(0.04,0.10)$. This comparison is consistent with the purpose of the sparsity correction. When layers within the same structural group have quite different sparsity levels, removing the layer-specific scale becomes more important before clustering.

\section{Application}\label{sec:application}

The 2024 BACI international trade data from CEPII \citep{gaulierZignago2010BACI} contain 226 economies and 68 product categories, each defining a network layer. Let $X_{ij}^{(l)}$ denote the trade value between economies $i$ and $j$ for product $l$, after combining the two trade directions. For pairs with positive total trade, we construct
\[
\cA_{ijl}
=\bbI\left\{
\frac{X_{ij}^{(l)}}{\sum_{h=1}^{L}X_{ij}^{(h)}}
>\frac{\sum_{a<b}X_{ab}^{(l)}}{\sum_{h=1}^{L}\sum_{a<b}X_{ab}^{(h)}}\right\},
\qquad i<j,
\]
and symmetrize $\cA$ with zero diagonal. Thus, an edge indicates that a product is more prominent in a bilateral trade relationship than it is in world trade overall.

We apply HINTS to $\cA$ with Tucker rank $(8,8,4)$, selected by cross-validation of edge prediction and an assessment of layer-partition stability. We first analyze the clustering of product layers.

\begin{figure}[htbp]
    \centering
    \includegraphics[width=0.9\linewidth]{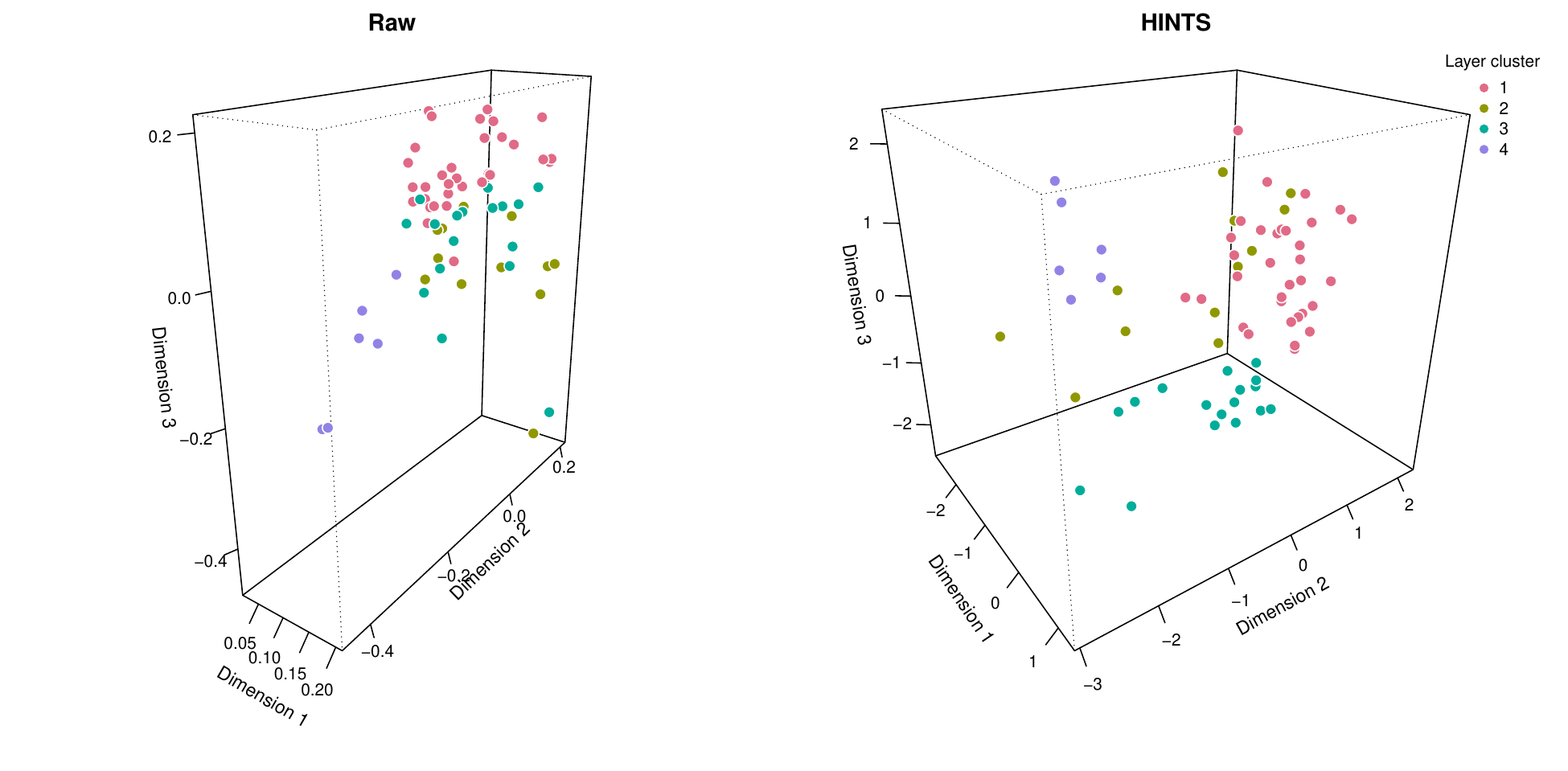}
    \caption{Raw and sparsity-corrected layer embeddings, colored by the product groups identified by HINTS.}
    \label{fig:BACI_embedding}
\end{figure}

Figure~\ref{fig:BACI_embedding} provides a geometric explanation. In the raw embedding, the layers exhibit a pronounced conic structure, and the four HINTS groups overlap substantially. Variation in sparsity mainly changes the radial magnitude of the embedding, so that differences in network density can dominate the distance between layers even when their underlying directions are similar. The sparsity correction removes this radial variation by rescaling the layer coordinates, retaining the directional information that characterizes their connectivity patterns. As a result, the four product groups become much more clearly separated in the corrected embedding.

The four product groups obtained by HINTS can be summarized as follows; the complete layer assignments are reported in the Supplementary Materials:

\begin{itemize}
    \item \textbf{Materials and basic manufactures}: inorganic chemicals, fertilizers, plastics, paper, textile materials, leather, iron and steel, and other base metals, etc.

    \item \textbf{Specialty chemicals and precision products}: organic chemicals, pharmaceuticals, cosmetics, photographic goods, aircraft, optical and medical instruments, clocks and watches, etc.

    \item \textbf{Machinery, transport, and other manufactures}: rubber and wood products, metal articles, mechanical and electrical machinery, railway and road vehicles, ships, furniture, toys, etc.

    \item \textbf{Apparel and accessories}: leather articles and luggage, apparel, footwear, headgear, basketware, etc.
\end{itemize}

The contrast with the raw embedding is informative. In the raw clustering, textile materials, railway equipment, ships, clocks and watches, and several metal products are placed in the same large low-density group, despite their rather different economic functions. After sparsity correction, railway equipment and ships are grouped with machinery and other transport products, while clocks and watches are grouped with precision products. Basketware is similarly moved from the large sparse group to the apparel-and-accessories group. These examples illustrate that the raw embedding can group economically distinct products simply because their network layers have similar sparsity, whereas HINTS produces groups that are more closely aligned with similarities in trading patterns.

The HINTS partition also separates materials from several related finished products. Textile materials and apparel belong to different groups, as do raw leather and leather articles, and iron and steel and fabricated metal articles. Thus, products that are closely related in a conventional industrial classification need not have similar networks of trading partners. Aircraft provide an interesting contrast. Although conventionally classified as transport equipment, their trading pattern is closer to those of specialty and precision products than to those of road vehicles, railway equipment, and ships. Taken together, these examples suggest that the organization revealed by HINTS reflects similarities in international trading relationships that are not captured by either marginal network density or broad product classifications alone.

We next examine the local community structures within the four product groups. The eigenvalue-ratio rule gives
$(K_1,K_2,K_3,K_4)=(4,2,2,2)$, and the complete clustering results are reported in the Supplementary Materials. The estimated communities vary considerably across product groups. For example, Germany and the United States belong to the more connected communities for specialty chemicals and precision products and for machinery and transport, whereas Bangladesh, Cambodia, Pakistan, and Viet Nam stand out in the apparel-and-accessories group. Such differences indicate that the position of an economy in the trade network can vary substantially across products. Imposing a common economy partition across all layers would obscure this product-specific structure.

\begin{figure}[htbp]
    \centering
    \includegraphics[width=\linewidth]{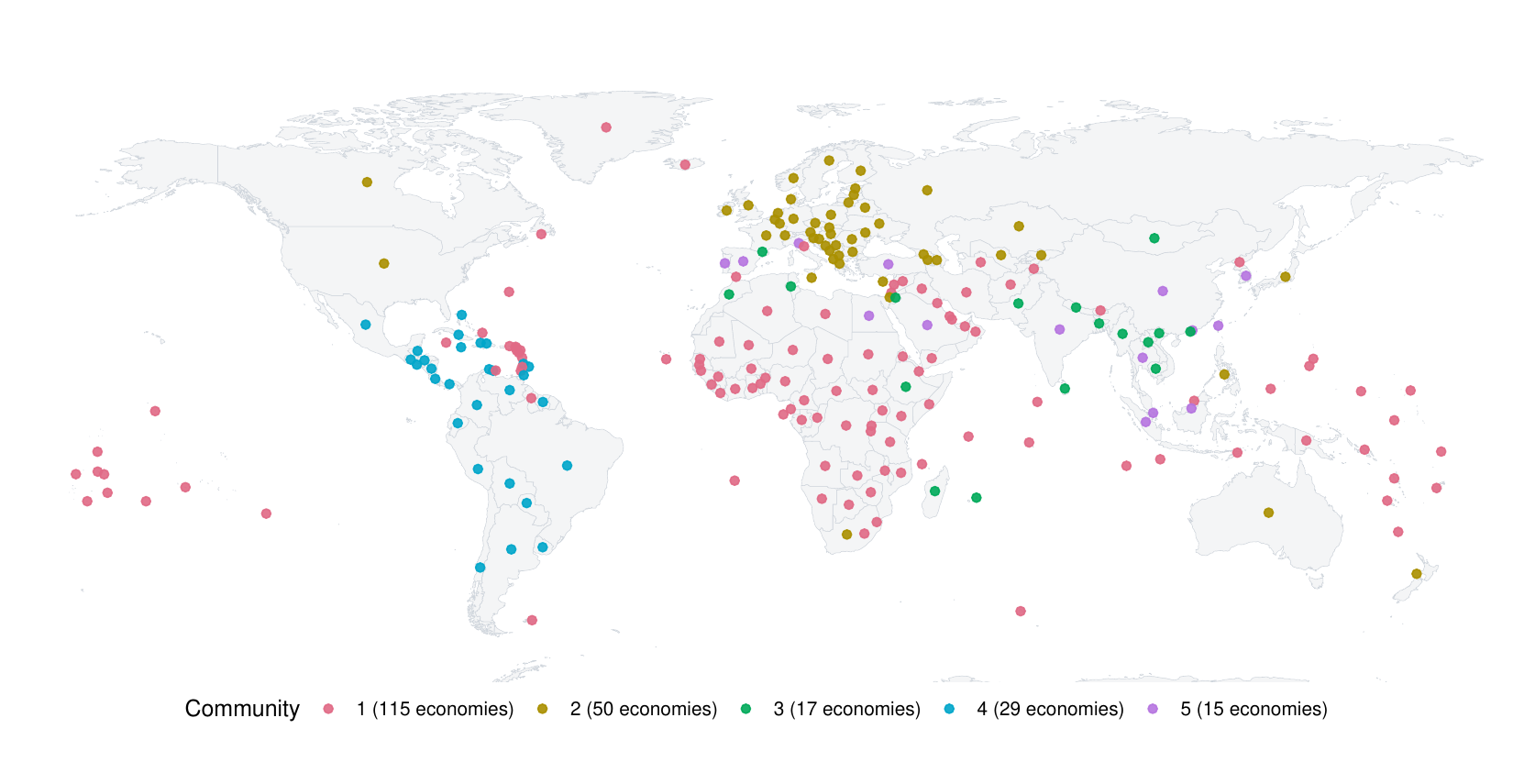}
    \caption{Global economy communities in the BACI trade networks. Geographic locations are used only for display and are not used in the clustering.}
    \label{fig:BACI_global}
\end{figure}

Combining the local community information yields five well-interpreted global communities, shown in Figure~\ref{fig:BACI_global}.

Several patterns are visible in Figure~\ref{fig:BACI_global}. Community 2 is centered on Europe and also contains the United States, Canada, Japan, and Australia, whereas Community 4 is concentrated in Latin America and the Caribbean. Since geographic information is not used in the analysis, these regional concentrations reflect similarities in the underlying trade networks rather than an imposed geographic partition.

More importantly, Communities 3 and 5 reveal two rather different forms of participation in international manufacturing trade. Community 3 includes Bangladesh, Cambodia, Pakistan, and Viet Nam and is distinguished by its strong apparel-related trade connections. Community 5 includes China, India, Indonesia, the Republic of Korea, Malaysia, Singapore, Thailand, together with several economies in Southern Europe, and exhibits strong connections across a much broader range of product groups. Thus, although economies in both communities are active in apparel-related trade, their broader trading structures are quite different: the former is characterized by a more concentrated apparel profile, while the latter is embedded simultaneously in materials, specialty products, machinery and transport, and apparel. The global communities therefore capture differences in both regional organization and product specialization that are not apparent from a single aggregate trade network.

Taken together, the analysis reveals trade structures at several complementary levels. The layer clustering separates products according to similarities in their trading networks rather than marginal sparsity, while the local and global economy communities uncover substantial heterogeneity in how countries participate across product markets. Such structure is relevant for understanding trade specialization, cross-sector dependence, and the uneven transmission of shocks across countries and industries.

\section{Conclusion}

Multi-layer networks contain information from multiple views, but this richness also brings analytical complexity. Pooling information across layers can improve estimation, yet differences in network sparsity need not correspond to differences in community structure. When these two sources of variation are mixed together, layer clustering may be driven by sparsity rather than by the patterns of interest. To account for this feature, SH-MMSBM provides a more realistic model for multi-layer network by separating layer-specific sparsity from the community structure shared by layers in the same group. 

Based on SH-MMSBM, HINTS is proposed for layer clustering and community detection. The tensor representation shows how layer sparsity affects the layer embedding and motivates a sparsity correction before clustering. The correction removes the layer-specific scale without estimating the individual sparsity parameters. After the layer groups are recovered, HINTS performs community detection within each group and combines the local results to obtain a global partition. In this way, information is pooled across related layers while group-specific community patterns are retained. 

The theoretical results show that substantial sparsity heterogeneity is compatible with consistent recovery. The allowable degree of heterogeneity depends on the information accumulated across nodes and layers. As heterogeneity increases, layer clustering becomes more difficult, but this effect can be offset by larger networks or more layers. In the homogeneous case, the extra cost disappears, and the resulting conditions and rates reduce to those of existing tensor-based methods. Numerical experiments show clear gains from HINTS when layer sparsity levels vary widely. The real-data applications also indicate that the local and global views can reveal different aspects of network organization. 

Beyond clustering, the estimated layer and node embeddings are useful for other downstream tasks, including change-point detection, link prediction, and network regression. Another natural extension is to incorporate node-level degree heterogeneity together with layer-wise sparsity heterogeneity. This would lead to a more flexible model, but would also raise new questions about identifiability and theory, and we leave this extension for future work.

\section*{Disclosure Statement}
    The authors declare that they have no conflict of interest.

\if1\anon
\section*{Acknowledgments}
    Gaorong Li and Xuefei Wang's research was supported by National Natural Science Foundation of China (Grant Number 12271046). Wenyang Zhang's research was supported by CPG2026-00039-FBA; CPG2025-00004-FBA; CPG2024-00004-FBA; SRG2023-00027-FBA.
\fi

\bibliography{ref.bib}

\end{document}